\documentclass[a4paper,11pt]{article}
\usepackage[a4paper, total={17cm, 25cm}]{geometry}
\usepackage{fancyhdr}
\usepackage{graphicx}
\usepackage{subcaption}
\fancypagestyle{plain}{
\fancyhf{}
\fancyhead[RH]{\thepage}
\fancyfoot[CF]{28th Texas Symposium on Relativistic Astrophysics \\ Geneva, Switzerland -- December 13-18, 2015}

}
\begin{document}
\title{\bf Angular Momentum Loss and Gravitational wave amplitudes for X-ray Binaries with a Neutron Star Component}

\author{
Tu\u{g}\c{c}e \.I\c{c}li$^1$ and Kadri Yakut$^1$ \\
$^1$ University of Ege, Faculty of Science, Department of Astronomy and Space Sciences, \.Izmir, Turkey}

\date{February 15, 2016}

\maketitle

\begin{abstract}
Binary systems with neutron stars and double degenerate systems are crucial objects to test current stellar evolution models and Einstein's general relativity.
In this study, we present angular momentum loss mechanism via gravitational radiation and magnetized stellar winds for
some selected systems with a neutron star. We calculated and plotted their time scales for angular momentum loss.
Gravitational wave amplitudes of binary systems with a neutron star components are also estimated and their detectability with a gravitational wave detector (LISA) has been plotted.
\end{abstract}


\section{Introduction}

Massive stars are progenitors of supernovae. At the end of their evolution they leave behind either neutron stars or black holes.
Stars follow different evolutionary stages up to this point of evolution \cite{Postnov2014}. These differences are mainly because of the initial masses of stars; especially for single stars. In binary stars, on the other hand, in addition to the initial mass; orbital period, mass ratio, and total mass are
crucial parameters that may affect their evolution. Evolution of X-ray binaries consisting of a compact object and a main-sequence star or a
giant star depends strongly on the variation of these parameters. X-ray binaries with low mass components are
classified as LMXBs while binaries with high mass components are classified as HMXBs.
Parameters of many binary systems that are members of LMXBs and HMXBs have been published in the literature.
In this study, X-ray binary systems consisting of a neutron star and a M/O star have been studied (NS+S). Both LMXB and HMXB are considered.

\section{Angular Momentum Loss in NS+S Systems}
The binary systems presented in this study contain both early and late type components.  Hence, mass loss via stellar activity that is observed in
 late type stars and mass loss through strong stellar wind, that is observed in early type stars, can be detected.
 In addition, mass transfer between the components takes place.  Mass loss/transfer is directly related to the angular
 momentum loss/transfer. In main-sequence stars angular momentum loss mechanisms via stellar wind are
 important (see \cite{Kawaler98}, \cite{Yakut05}, \cite{Yakut08}) while in cataclysmic variables and in
 other X-ray binaries mass loss via stellar activity and angular momentum loss via gravitational radiation is important
 (see \cite{Andronov03}, \cite{Kalomeni10} and references therein).

If the components are close to each other, in other words if the orbital period is short, angular momentum loss via
gravitational radiation might be important.  General relativity predicts energy loss because of the perturbation in space-time of binary stars,
which releases gravitational waves \cite{EInstein18}. Recently, gravitational waves are observed by LIGO team \cite{LIGO2016a}.

In this project, the orbital and physical parameters of 33 systems have been collected. Collecting the parameters we
carefully examined all the studies and selected the best observational results. The orbital period vary from 0.2 to 1236 days.
The mass of the components varies between 0.24 to 58 solar masses. Some of the systems have late type components; therefore,
in addition to angular momentum loss via mass loss, angular momentum loss through gravitational radiation is also important.
Angular momentum loss via stellar winds and gravitational radiation has been studied previously by \cite{Yakut08} and the formulae are given in Eq. (1) and Eq. (2). Yakut et al. \cite{Yakut08} calculated angular momentum loss time scale via stellar wind and gravitational radiation (Eq.~1 and Eq.~2).

\begin{equation}
\tau_{MSW} = 14 (\frac{M}{M_\odot})^{2/3} (\frac{R_2}{R_\odot})^{-4}(\frac{P}{\rm{day}})^{10/3}(1+q)^{-1}(1-e^2)^{1/2} \rm{Gyr}.\label{Eq-timescale-msw}
\end{equation}

\begin{equation}
\tau_{GR} = 376.4q^{-1}(1+q)^2M^{-5/3}P^{8/3} (1-e^2)^{3/2}(1+\frac{7}{4}e^2)^{-1}  \textrm{Gyr}. \label{Eq-timescale-gr}
\end{equation}

We have calculated angular momentum loss time scales for x-ray binaries with neutron star. We plotted the results in Fig. 1a and Fig. 1b.

\begin{figure}
\includegraphics[width=.5\textwidth]{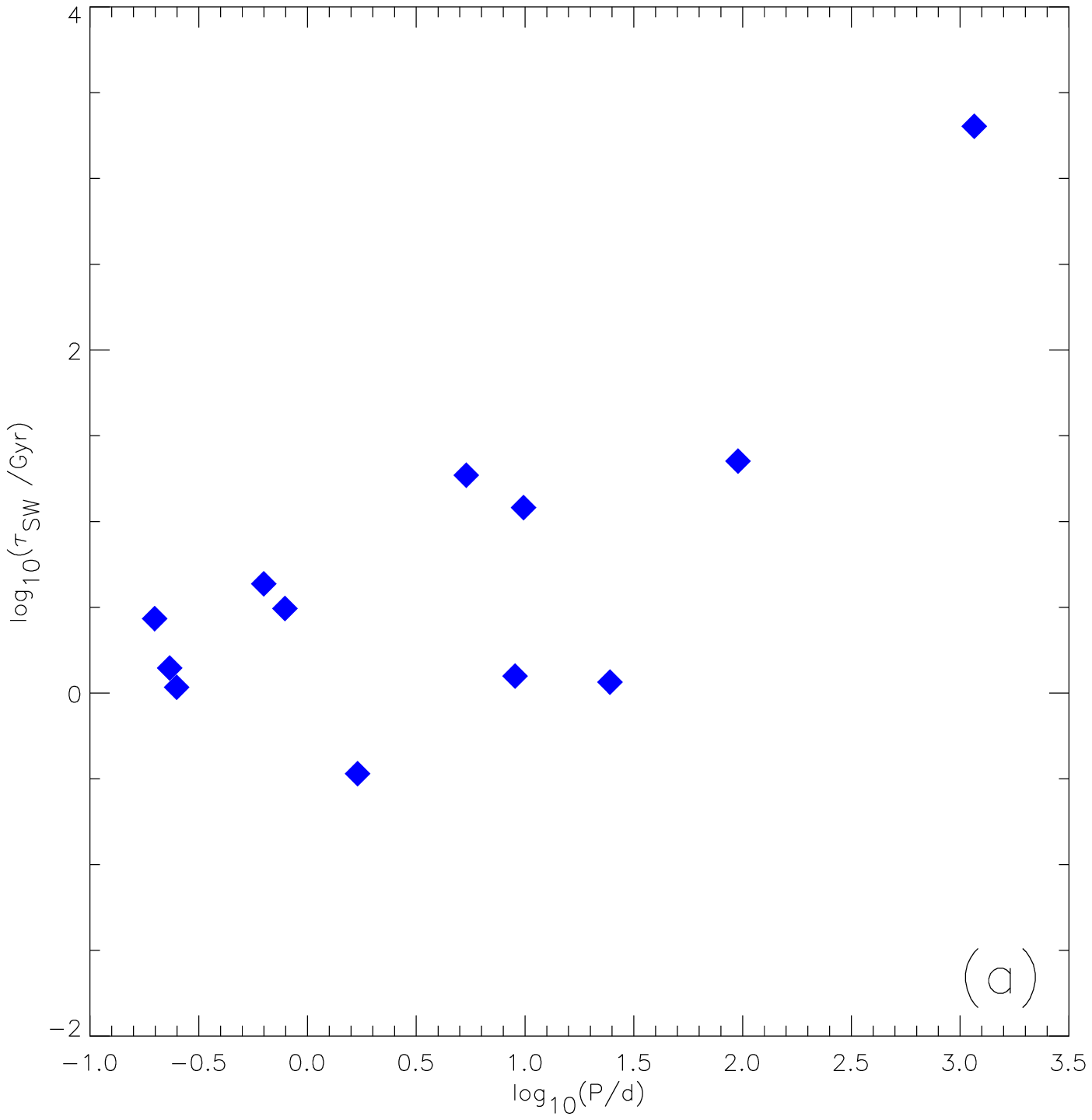}
\includegraphics[width=.5\textwidth]{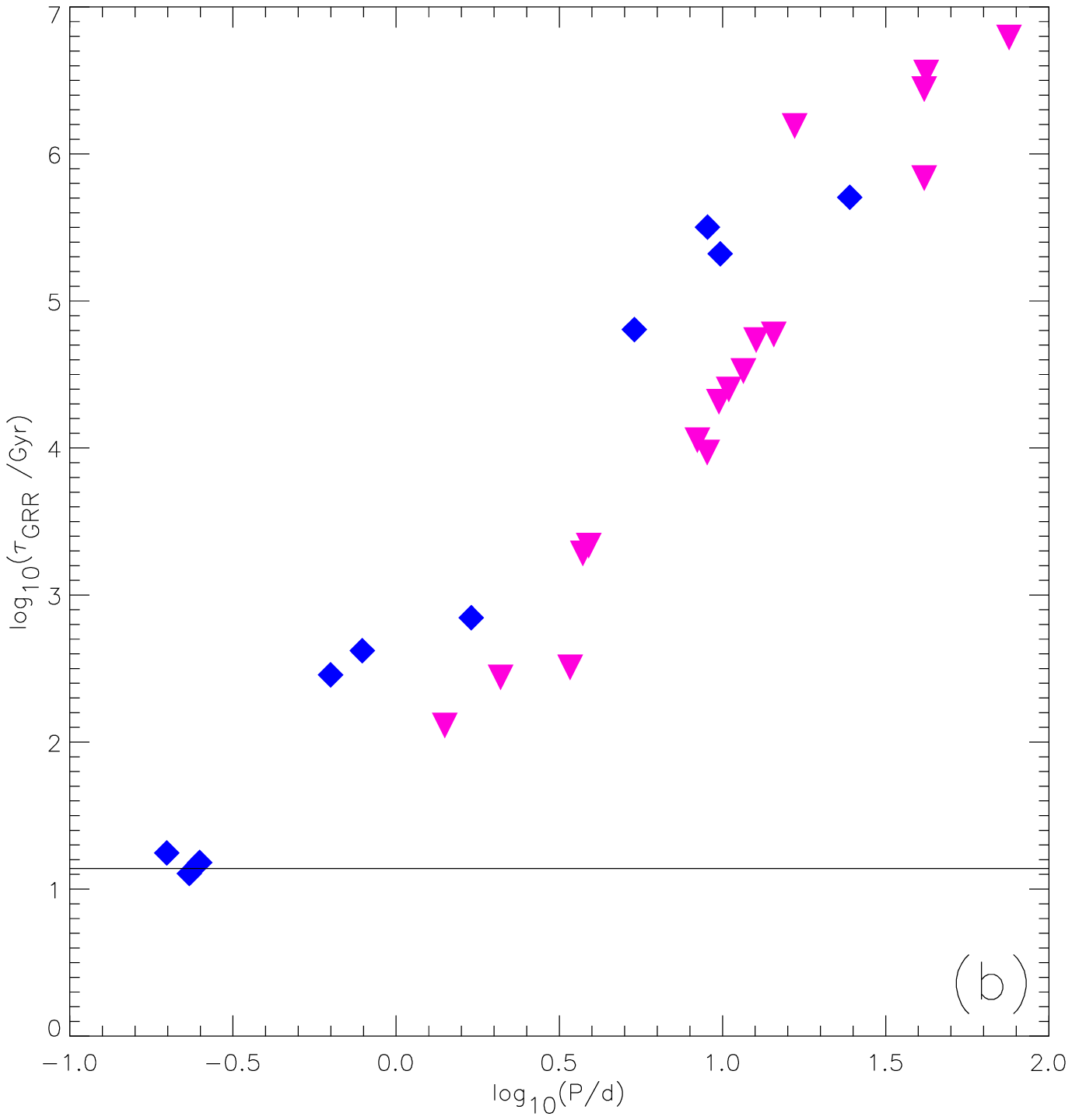}
\caption{(a) Plot of  $\log \tau_{\textrm{MSW}}$ {\it vs.} $\log$P and (b) $\log \tau_{\textrm{GR}}$ {\it vs.} $\log$P for binary systems with neutron star.}\label{fig1}
\end{figure}

\begin{figure}
\includegraphics[width=.5\textwidth]{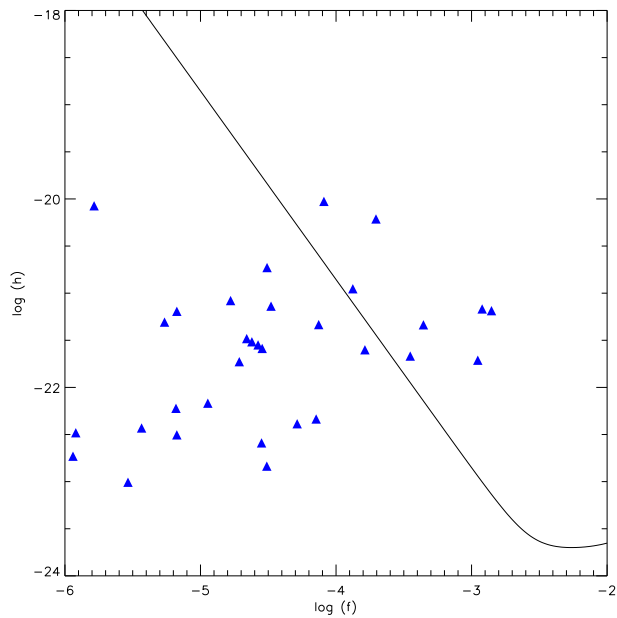}
\caption{Plot of  $\log h$ {\it vs.} $\log f$ for binary systems with neutron star components. Continuous line shows the expected sensitivity of LISA.}\label{fig3}
\end{figure}

Observations of gravitational waves (Abbott et al. 2016) increased the importance of compact binaries.
Binary systems with neutron systems could be the progenitor of NS+NS or NS+BH systems and most important candidates of gravitational-wave sources.
K{\"o}se and Yakut \cite{OY2012} calculated the GW amplitude (h) following the linearized Einstein field equation (Eq. 3).
We calculated h amplitude for the collated NS+S systems and plotted them in Fig.~(2).

\begin{equation}
h \simeq 2.5 \times 10^{-22} {M_{1}M_{2}}{M^{-1/3}D^{-1}}f^{2/3} \label{amplitude}
\end{equation}

\paragraph{}
\paragraph{}
\textbf{Acknowledgments}\\
This study was supported by the Turkish Scientific and Research Council (T\"UB\.ITAK 113F097).
The current study is a part of MSc thesis by T. \.I\c{c}li.
\paragraph{}

\end{document}